\documentstyle{article}
\begin{document}
\title{\Large \bf Adsorption on a periodically corrugated substrate}
\author{{\large K.Rejmer and M.Napi\'orkowski } \\Instytut Fizyki 
Teoretycznej, 
Uniwersytet Warszawski \\00 681 Warszawa, Ho\.za 69, Poland \vspace*{5mm}}
\maketitle
\begin{abstract}{Abstract. Mean field analysis of the effective interfacial 
Hamiltonian shows that with increasing temperature the adsorption on a 
periodically corrugated substrate can proceed in two steps: 
first, there is the filling transition in which the depressions of the 
substrate become partially or completely filled; then there is the wetting 
transition 
at which the substrate as a whole becomes covered with a macroscopically 
thick wetting layer. The actual order and location of both transitions 
are related to the wetting properties of the corresponding planar substrate 
and to the form of corrugation. Certain morphological properties of the 
liquid-vapor interface in the case of a saw-like corrugated substrate are 
discussed analytically.} \\ 

\noindent PACS numbers : 68.45.Gd, 68.35.Md, 68.35.Rh 
\end{abstract}
\newpage

\centerline{\bf {I. Introduction}}
\renewcommand{\theequation}{1.\arabic{equation}} 
\setcounter{equation}{0}
\vspace*{0.5cm}

Although the wetting of homogeneous and planar substrates is 
nowadays a relatively well understood phenomenon [1-3] its counterpart 
corresponding to the adsorption on corrugated substrates still poses many 
questions which remain unanswered in spite of much recent experimental and 
theoretical work [4-19] devoted to these systems. Different apects of wetting 
on a corrugated substrate, e.g. the order of the wetting transition, the 
location of the wetting point on the thermodynamic phase diagram, and the 
structure of the emerging liquid-vapor interface  have been usually  
refered to the wetting properties of the corresponding planar substrate 
which indeed serves as the natural reference system. However, recent work 
on the special type of non-planar substrate - the infinitely extended 
wedge-like substrate [20-23] - points at a novel type of 
transition characteristic for this system. This is so-called filling 
transition which - in our opinion - becomes also relevant in the case of 
adsorption on a periodically corrugated substrate. It takes place at the 
bulk liquid-vapor coexistence and at a temperature that is lower than the 
wetting temperature of the corresponding planar substrate.  
In the course of the filling transition the width od the wetting film becomes 
macroscopically thick in the center of the wedge and it remains thin 
far away from the wedge center; there the interface locally interacts 
with the planar substrate only which at the filling transition remains 
nonwet. 

Our mean-field analysis of adsorption on a periodically corrugated substrate 
aims at showing that the filling transition mentioned above can be also 
relevant in this case and precedes the wetting transition [19]. Thus two 
kinds of transitions, i.e. the filling and the wetting transition can be 
present in the scenario of adsorption on the corrugated substrate. 
 
In Chapter II we introduce the system and present the thermodynamic analysis 
of the problem. It shows which properties of the substrate's shape are 
relevant for the filling transition. We also comment on the 
phenomenological Wenzel's law [24,25] locating the wetting temperature for a 
corrugated substrate. Chapter III contains the analysis of the adsorption 
in the case when the corrugation of the substrate has the special saw-like 
form. This analysis is based on rather general  
considerations using the concept of the effective interfacial Hamiltonian. 
They allow us to extract information not only about the order of the filling 
and the wetting transitions but also point at certain structural properties of 
the liquid-vapor interface. Chapter IV contains analytic discussion of the 
above issues. In this case the shape of the interface can be obtained 
explicitely and its scaling properties in different temperature regimes can 
be transparently presented. In Chapter V we summarize our results. \\

\centerline{\bf {II. Thermodynamic description}}
\renewcommand{\theequation}{2.\arabic{equation}} 
\setcounter{equation}{0}
\vspace{0.5cm}

We consider a substrate which is periodically corrugated in the x-direction 
and 
translationally invariant in the y-direction. It's shape is described by 
the function $z = b(x) $ with period $2a$, i.e. $b(-a)=b(a)$. Moreover, we 
assume that $b(-x)=b(x)$ and that  $b(x)$ is monotonically increasing for 
$0 <x < a$;  $b(x)$ has the minimum at $x=0$ and the maximum at $x=a$. The 
space above the substrate is occupied by the inhomogeneous fluid and the 
thermodynamic conditions are chosen such that for $z \rightarrow \infty$ the 
bulk vapor is the stable thermodynamic phase. Close to the substrate, due to 
its preference for a liquid phase a liquidlike layer is formed which separates  
the vapor from the substrate. The knowledge of the shape of the liquid-vapor 
interface $z=f(x)$ and the width of the liquid-like layer $l(x)=f(x)-b(x)$ as 
function of the thermodynamic state of the system allows one to discuss the 
possible filling and the wetting transition taking place in this system.  
We do not put any constraints on the amount of liquid adsorbed 
on the substrate. Such contraints might lead to the interfacial configurations 
which break the substrate symmetry [26]; this is not the case in our 
analysis. 

In this Chapter we analyse the problem thermodynamically by considering 
macroscopic configurations of the interface and evaluating the 
corresponding surface free energies.  In order to make the notation consistent 
with that employed later within the effective Hamiltonian approach 
(Chapters III and IV) we denote the vapor phase and the liquid phase as 
phases $\alpha$ and $\beta$, respectively. 

We assume that the substrate imposes its periodicity on the allowed 
interfacial configurations and thus our analysis is reduced to a single 
subtrate's depression extending on the segment $[-a, a]$. Moreover, 
due to the substrate's symmetry $b(x)=b(-x)$ it is enough to concentrate on 
the segment $[0,a]$. One takes into account two types of 
competing interfacial configurations: the first corresponds to the 
partial filling of the depression, Fig.1a, and the second corresponds to the 
interface located above or exactly at the top of the substrate, i.e. 
$f(x) \geq b(a)$, see Fig.1b. 
The free energy of the depression filled completely with the phase $\alpha$  
serves as the reference point. Then the excess surface free energies 
corresponding to the above two cases  have the following form 
\begin{eqnarray}
\Delta F_{1}(x_{1}) = 2 \,x_1 \,\sigma_{\alpha\beta} + L(x_1)\,
(\sigma_{w\beta} - \sigma_{w\alpha})\, =\nonumber \\ 
=\,\sigma_{\alpha\beta}\,[2x_1 - L(x_1)\,\cos\theta] \,\,\,,
\end{eqnarray}
\begin{eqnarray}
\Delta F_{2}\, =\, \sigma_{\alpha\beta}\,[2a - L(a)\,
\cos\theta] \,\,\,,
\end{eqnarray}
respectively. $x_{1}$ denotes the value of the abscissa at which the planar 
segment of the 
$\alpha$-$\beta$ interface makes the contact with the substrate, i.e. 
$f(x_1)=b(x_1)$.  $\sigma_{\alpha\beta}$, $\sigma_{w\alpha}$, 
$\sigma_{w\beta}$ are the $\alpha$-$\beta$, substrate-phase $\alpha$, 
substrate-phase $\beta$ surface tensions, and 
\begin{eqnarray}
L(x_1)=
\int\limits_{-x_1}^{x_1}\,dx\,\sqrt{1+b_{x}^{2}(x)}
\end{eqnarray}
is the length of the substrate-phase $\beta$ interface. In Eqs.(2.1-2.2) 
the Young equation has been used and $\theta$ denotes the planar substrate 
contact angle. \\
The equilibrium interfacial configuration corresponds to the value 
$ \bar{x}_{1}$ of the contact point abscissa which minimizes the excess 
free energy. It solves the following equation
\begin{eqnarray}
0\,=\,\Delta F_{1,x}(\bar{x}_1) \,=\,  2 \sigma_{\alpha\beta}\, 
\left(1-\sqrt{1+b_{x}^{2}(\bar{x}_1)}\,\cos\theta \right) \,= \nonumber \\  
 2 \sigma_{\alpha\beta}\,
\left(1-\frac{\cos\theta}{\cos \psi(\bar{x}_1)}\right)\,\,,
\end{eqnarray}
where $\psi(\bar{x}_1)$ denotes the actual contact angle on the corrugated 
substrate 
shown in Fig.1a. Thus the surface free energy has an extremum for such value 
of $\bar {x}_1$ that the corresponding contact angle $\psi(\bar{x}_1)$ 
becomes equal to the contact angle $\theta$ on the planar substrate. 
The actual minimum is located by inspecting the second derivative of the 
excess free energy
\begin{eqnarray}
\Delta F_{1,xx}(\bar{x}_{1})\,=\,-2\,\sigma_{\alpha\beta}\,
\frac{b_{x}(\bar{x}_1)\,b_{xx}(\bar{x}_1)}{\sqrt{1+b_{x}^{2}(\bar{x}_{1})}}
\,\cos\theta\,\,.
\end{eqnarray}
For substrates under consideration one has $b_{x} > 0$ for $0 < x < a$ and 
the sign of $\Delta F_{1,xx}(\bar{x}_{1})$ is determined by the sign of 
$b_{xx}(\bar{x}_{1})$, i.e. by the curvature of the substrate. 
$\bar{x}_1$ corresponds to the excess free energy minimum if the substrate is 
convex at the corresponding point $\bar{x}_1$ (i.e. $b_{xx}(\bar{x}_1) < 0$ 
and $b(x)$ is concave at $\bar{x}_{1}$).  
On the other hand if the substrate is concave at  $\bar{x}_1$  
(i.e. $b_{xx}(\bar{x}_1) > 0$) then the excess free energy has maximum. 
For an arbitrary substrate shape $b(x)$ with several inflection points 
there may be many competing equilibrium states leading to a rather 
complicated phase diagram. 

For periodic substrates with a single inflection point $x_{inf} \in [0,a]$ 
the  situation is more transparent. An example of such 
a substrate is given by $b(x)= a \,(1-\cos(x\pi/a))$ for which 
one has $b_{x}(0)=b_{x}(a)=0$ and $x_{inf}=a/2$. For appropriate range of 
values 
of the contact angle $\theta$ the excess free energy has both maximum and 
minimum located 
at $x_{max}$ and $x_{min}$, respectively. Upon increasing the temperature 
$\theta$ decreases, $x_{max}$ decreases, and $x_{min}$ increases. The 
limiting 
value $\theta_{0}$ such that for $\theta < \theta_{0}$ these two extrema of 
the excess free energy exist is determined by the shape of the substrate: 
$\theta_{0} \,=\,\psi(x_{inf})$. 
For temperatures below $T_{0}$ such that $\theta(T_{0})=\theta_{0}$ the 
depression remains filled with the bulk phase $\alpha$. This case is called 
the empty depression and  $\bar{x}_{1}=0$ . For $T > T_{0}$  the depression  
may be partially filled with (unstable in the bulk) phase 
$\beta$; $\bar{x}_{1}=x_{min} \in [x_{inf}, a]$. The equilibrium 
configuration is selected by the excess free energy balance, Fig.2. 
The transition from 
the empty to the partially filled configuration which takes place at 
$T_f$ is first-order. The equilibrium value $\bar{x}_{1}$ changes 
discontinuously from $0$ to $\bar{x}_{1f}$ which is the smallest value 
of $x_{min}$ such that $\Delta F_{1}(\bar{x}_{1f})\leq 0$. We call it the 
filling transition. The corresponding value $\theta_f$ of the contact angle 
is given by 
\begin{eqnarray}
\cos\theta_{f}\,=\,\frac{2\,\bar{x}_{1f}}{L(\bar{x}_{1f})}\,\,.
\end{eqnarray}
Upon further increase of 
the temperature $x_{min}$ increases towards $a$; this limiting value of 
$\bar{x}_{1}$ at 
which the whole depression is filled with the phase $\beta$ is achieved 
for  $\theta = 0$, i.e. at the planar substrate wetting temperature $T_{w}$. 
Still further increase of the temperature does not change the excess free 
energy balance and the whole substrate remains covered by the $\beta$-like 
layer of growing width. However, the analysis of this further growth is 
beyond the present thermodynamic description. 

The above analysis shows that the very existence of a nontrivial filling 
transition depends substantially on the substrate's shape.  
If each depression is strictly concave (i.e. $b_{xx}(x) > 0,\, x \in (-a,a)$)  
then the thermodynamic argument points at non-existence of the filling 
transition leading to a {\it partially} filled depression. For example,  
when the periodic piecewise concave substrate consists of the arcs of 
circles of radius $R$ then 
\begin{eqnarray}
\Delta F(x_1)\,=\,2\,R\,\sigma_{\alpha\beta}\,\left[x_{1}/R\,-\,
\arcsin(x_{1}/R)\,\cos{\theta}\right]\,\,\,,
\end{eqnarray}
see Fig.3. Upon increasing the temperature the depression remains empty 
until the contact angle reaches the value $\theta_{f}$ such that 
$\cos{\theta_{f}}=\sin{\delta} / \delta$ at which the whole depression 
becomes filled with the $\beta$-phase and $\bar{x}_1=a$. In this case 
the circle's arc forming the depression is tangent to the $x$-axis at 
$x=0$ and  $\pi - 2\delta$ denotes the angle between the two arcs meeting 
at $x=a$. On the contrary, for piecewise strictly convex substrate 
(i.e. $b_{xx}(x) < 0$ for $ x \in [-a,0)$ and $x \in (0,a]$) 
the filling of the depression starts at its center and proceeds continuously 
upon increasing the temperature up to the point $x_{1}=a$. 
For example, when the periodic piecewise convex substrate consists of the 
arcs of circles of radius $R$ then  
\begin{eqnarray}
\Delta F(x_1)\,=\,2\,R\,\sigma_{\alpha\beta}\,\left[x_{1}/R\,-\,
\left(\delta\,-\,\arcsin(\sin\delta-x_{1}/R)\right)\cos{\theta} 
\right]\,\,\,,
\end{eqnarray}
see Fig.4. In this case the tangent to each of the two circle arcs forming 
the depression at $x \in [-a,a]$ is horizontal at $x=\pm a$ and 
$\pi - 2\delta$ denotes the angle between the two arcs meeting  at $x=0$. \\
In this case the depression becomes completely filled at $\theta = 0$, i.e. 
at $T=T_{w}$. 

The above macroscopic discussion also sheds light on the phenomenological 
Wenzel's criterion [24,25] locating the wetting temperature $T_r$ for a 
rough substrate. Accordingly, if one compares two 
interfacial configurations corresponding to the depression either 
completely filled with the $\alpha$-phase (the so-called empty depression) 
or completely filled with the $\beta$-phase then equating their free 
energies leads to the following equation for $T_r$ 
\begin{eqnarray}
\cos\theta(T_{r})\,=\,\frac{2\,a}{L(a)}\,\,.
\end{eqnarray}
The inverse of the r.h.s. of the above equation is the so-called roughness 
factor which measures the ratio of the actual corrugated surface area to 
its projection on the plane. Eq.(2.9) determines - according to the Wenzel's  
criterion - the wetting temperature of the rough substrate as 
compared to the planar substrate case for which $\theta(T_{w})=0$. However, 
this phenomelogical criterion disregards the partially filled 
configurations and the transitions leading to them which may actually 
preempty the transition to a completely filled depression. In addition, this 
criterion does not distinguish transitions to the completely filled 
depression and those to the state in which the interface is removed 
macroscopically 
far away from the substrate as a whole. On the other hand if one looks at 
the filling transition for an infinitely extended wedge - which is the 
special limiting case of the substrates we consider here - then the filling 
transition temperature is correctly given by the Wenzel's rule. 
The same holds true for the special kind of concave substrate whose 
free energy is described by Eqs.(2.7). 

In the case of the periodic saw-like substrate, i.e. 
$b(x)\,=\,|x|\cot{\varphi}$, where $\varphi$ denotes half of the saw 
opening angle the above thermodynamic arguments point at 
the problems which need to be resolved at the more microscopic level. 
The excess free energy 
\begin{eqnarray} 
\Delta F(x_1)\,=\,2\,\sigma_{\alpha\beta}\,x_{1}\,
\left(1-\cos{\theta}/\sin{\varphi}\right) 
\end{eqnarray}
is linear function of $x_{1}$. The positive values of the coefficient 
$(1-\cos{\theta}/\sin{\varphi})$ 
correspond to the empty depression and the negative values to the completely 
filled case. At the transition temperature $T_{\varphi}$ such that 
$\cos{\theta}(T_{\varphi})=\sin{\varphi}$ no $\bar{x}_1$-value is 
distinguished 
by the present macroscopic argument. This corresponds to a degenerate 
case in which each configuration with $\bar{x}_1 \in [0,a]$ has the same 
value of the surface free energy. One certainly needs a more microscopic 
approach to discuss this case and to distinguish between the different 
interfacial configurations with the same surface energy. This approach, 
in addition to the surface contributions to the free energy should involve 
the analysis of the line contributions as well. The next chapter is devoted 
to this problem.  \\ 

\centerline{\bf {III. The interfacial configurations}}
\renewcommand{\theequation}{3.\arabic{equation}} 
\setcounter{equation}{0}
\vspace{0.5cm}

In this chapter we analyze the interfacial configurations in the 
presence of the periodic saw-shaped substrate introduced above. We restrict 
our analysis to a single section $x\in[-a,a]$ in which the substrate is 
described by $z=b(x)=|x|\cot{\varphi}$. Our mean-field approach is based on 
the effective Hamiltonian description which has been rather 
succesfully employed in discussing various interfacial problems 
[2,3,15-19,22,23].  \\ 

\centerline{\bf {IIIa. The effective Hamiltonian}}
\renewcommand{\theequation}{3.\arabic{equation}} 
\setcounter{equation}{0}
\vspace{0.3cm}

The effective Hamiltonian has the following form
\begin{eqnarray} 
{\cal H}[f]=\int \limits_{-a}^{a}dx\left\{ \frac{\sigma_{\alpha\beta}}{2}\,
\left[\left(\frac{df(x)}{dx}\right)^2\,-\,\cot^{2}{\varphi}\right]\,+\,
\frac{V(l(x))}{\sin{\varphi}}\,\right\}\,,
\end{eqnarray}
where the film thickness $l(x)=f(x)-|x|\cot\varphi$ is measured along the 
z-axis. The form of the effective potential $V(l)$  which desribes the 
interaction of the interface with the substrate [1-3,15-19,22] will be 
specified later depending on 
the order of the wetting transition on the planar substrate that we want to 
refer to. The above form of the effective Hamiltonian is valid for not too 
rough substrate, i.e. for $\varphi$ close to $\pi/2$ or 
$\cos^2{\varphi} \ll 1$. 

The equilibrium interfacial configuration $\bar f$ minimizes ${\cal H}[f]$ 
and solves the equation
\begin{eqnarray} 
{\sin}{\varphi}\,\sigma_{\alpha\beta}\,\frac{d^2\bar{f}(x)}{dx^2}\,=
\,V'(\bar{l}(x))
\end{eqnarray}
supplemented by the boundary conditions $\bar{f}'(0)\,=\,\bar{f}'(a)\,=0$\,. 
(As before, due to the symmetry of the problem we restrict our analysis to 
the segment $[0,a]$.)  
Integrating Eq.(3.2) one obtains
\begin{eqnarray} 
\frac{\sigma_{\alpha\beta}}{2}\,\sin\varphi
\left[\left(\frac{d\bar{l}}{dx}\right)^2\,-\,
\left.\left(\frac{d\bar{l}}{dx}\right)^2\right|_{x=0}\right]
\,=\,V(\bar{l}(x))\,-\,V_{1}\,\,,
\end{eqnarray}
where $\bar{l}_{1}=\bar{l}(0)$ and $V_{1}=V(\bar{l}_{1})$. Since 
$\bar{l}'(0^{+})=\bar{l}'(a^{-})=-\cot{\varphi}$ it follows from 
Eq.(3.3) that 
$V(\bar{l}_{1})=V(\bar{l}_{2})=V_{1}$, where $\bar{l}_{2}=\bar{l}(a)$. 
 
The equilibrium solution of Eq.(3.3) ${\bar l}(x)$ fulfills the 
constraint 
\begin{eqnarray}
a\,=\,\sqrt{\frac{\sin \varphi}{2}}
\int \limits_{{\bar l}_2}^{{\bar l}_1} dl\left[\sigma^{-1}_{\alpha \beta}
\,\Delta V(l)\,-\,\sigma^{-1}_{\alpha \beta}\,
\Delta V_1 \,+\,v(\varphi)\right]^{-\frac{1}{2}}\,\,\,,
\end{eqnarray}
where $\Delta V(l)=V(l)-V(l_{\pi})$, $\Delta V_{1}=V_{1}-V(l_{\pi})$, 
and $v(\varphi)\,=\,\frac{1}{2}\,\cot^2\varphi \sin \varphi$. $l_{\pi}$ is 
the equilibrium thickness of the wetting layer on the planar substrate and 
corresponds to the global 
minimum of $V(l)$. Eq.(3.4) together with the condition 
$V(\bar{l}_{1})=V(\bar{l}_{2})=V_{1}$ allows one to find the  pair 
$(\bar{l}_{1}, \bar{l}_{2})$ which characterizes the equilibrium interfacial 
configuration, see Figs 5 and 6. \\

\centerline{\bf {IIIb. The free energy decomposition}}
\renewcommand{\theequation}{3.\arabic{equation}}
\vspace{0.5cm}

The free energy of the system is obtained from the Hamiltonian in Eq.(3.1) 
evaluated at the equilibrium interfacial configuration ${\bar f}$ :
\begin{eqnarray}
{\cal H}[{\bar f}]\,=\,\frac{2\,a\,\sigma_{\alpha \beta}}{\sin \varphi}
\left(\sigma^{-1}_{\alpha \beta}\Delta V_1\,-\,v(\varphi)\right)\,+
\nonumber\\
2\,\sigma_{\alpha \beta}\,\sqrt{\frac{2}{\sin \varphi}}
\int \limits_{{\bar l}_2}^{{\bar l}_1} dl
\left\{\sqrt{\sigma^{-1}_{\alpha \beta}\Delta V(l)\,-\,
\sigma^{-1}_{\alpha \beta}\Delta V_1\,+\,v(\varphi)}\,-\,\sqrt
{v(\varphi)}\right\}\,\,\,. 
\end{eqnarray}
Our purpose is to extract the line and the surface contributions to
the free energy in Eq.(3.5) in the limit of large $a$ and to discuss their 
nonanalyticities corresponding to the transitions taking place in the system. 
This can be - at least partially - achieved via the graphical analyses 
presented below. 

Figures 5 and 6 display the potential difference 
$\sigma_{\alpha \beta}^{-1}\,\Delta V (l)$ in the case of the first-order 
(Figs 5a, 6a) and continuous (Figs 5b, 6b) transitions. Figures 5a,b 
correspond to $\cos \Theta\,<\,\sin \varphi$, 
i.e. $T\,<\,T_{\varphi}$, where $\theta(T_{\varphi})\,=\,\frac{\pi}{2}
\,-\,\varphi$. $T_{\varphi}$ is thus identified as the filling
temperature for a wedge with the opening angle $2\varphi$ [22]. Figures 
6a,b correspond to $T\,>\,T_{\varphi}$ . The dotted line represents 
$\sigma_{\alpha \beta}^{-1}\,\Delta V_1$; its intersections with the 
plot of 
$\sigma_{\alpha \beta}^{-1}\,\Delta V(l)$ determine, together with
Eq.(3.4), the equilibrium values of $l_1$ and $l_2$. 

We start with the first-order transition case for 
$T < T_{\varphi}$ (Fig.5a) and consider the special limit 
$a\,\rightarrow \,\infty$  in which the integral on the r.h.s. of Eq.(3.4) 
diverges. This divergence occurs either because the integrand diverges or 
because the upper limit ${\bar l}_1$ of integration in Eq.(3.4) diverges. 
For $T < T_{\varphi}$ only the first possibility can be realized; it happens 
for  $\sigma_{\alpha \beta}^{-1}\,
\Delta V_1\,\nearrow \,v(\varphi)$. Then the second term of the r.h.s. of 
Eq.(3.5) remains finite. The first term is a product of two factors; the 
first factor grows linearly with $a$ and the second one decreases to zero. 
Thus the first term gives no contribution to the surface free energy; both 
terms contribute to the line free energy. It is clear from Fig.5b that 
similar behavior will be observed for the second-order transition as well. 

The situation  changes  for $T>T_{\varphi}$, Figs 6a,b. In the case of the 
first-order transition (Fig.6a) there is a competition between two solutions  
corresponding to $({\bar l}_1,\,{\bar l}_2)$ and  
$({\tilde l}_1,\,{\tilde l}_2)$. 
Note that since both solutions fulfill the constraint given in 
Eq.(3.4) the values of $\Delta V({\bar l}_1)\,=\,\Delta V({\bar l}_2)$ and 
$\Delta V({\tilde l}_1)\,=\,\Delta V({\tilde l}_2)$ denoted as 
$\Delta V_1$ and $\Delta \tilde{V}_1$, respectively are different.   
The solution $({\tilde l}_1,{\tilde l}_2)$ and the 
corresponding free energy share the features discussed previously 
for the case $T\,\leq\,T_{\varphi}$ and $a\,\rightarrow \,\infty$. In 
particular, the free energy contains only the line contribution. The solution 
$({\bar l}_1,{\bar l}_2)$ 
and the corresponding free energy behave differently; in this case 
the divergence of the integral on the r.h.s. of Eq.(3.4) is provided by 
the diverging upper limit of integration. The first term on the r.h.s. of 
Eq.(3.5) can be rewritten (up to the $(\pi/2-\varphi)^2$ terms) as 
\begin{eqnarray}
2\,a\,\sigma_{\alpha \beta}\left\{\frac{1}{\sin \varphi}
\left[\sigma^{-1}_{\alpha \beta}\Delta V_1\,-\,
(1\,-\,\cos \theta)\right]
\,+\,\left[1\,-\,\cos \theta/\sin{\varphi}\right]\right\}\,\,\,.
\end{eqnarray}
Thus for large $a$ the first term in the above expression ceases to be 
linear in $a$ while the second term remains negative and proportional 
to $a$. The integrand in the second term on the r.h.s. of Eq.(3.5) also 
remains finite in this limit. For $T>T_{\varphi}$ 
the free energy corresponding to the solution $({\bar l}_1,\,{\bar l}_2)$  
contains an additional {\it negative} surface contribution. Due to this term 
the solution $({\bar l}_1,\,{\bar l}_2)$ corresponds to the actual equilibrium 
configuration. Since both solutions $({\bar l}_1,\,{\bar l}_2)$ and 
$({\tilde l}_1,\,{\tilde l}_2)$ correspond to the same value of the 
parameter $a$ one has $\Delta \tilde{V}_{1} > \Delta {V}_{1}$. For certain 
range of not too small $a$-values the difference 
${\cal{H}}[\tilde{f}] - {\cal{H}}[\bar{f}]$ may change sign depending on 
the temperature. This  marks the existence of the filling temperature  
$T_{\varphi}(a)$ such that $T_{\varphi}(a) > T_{\varphi}(\infty)=T_{\varphi}$ 
at which the transition between the two solutions takes place. The 
difference $T_{\varphi}(a)-T_{\varphi}$ decreases for $a \rightarrow \infty$; 
this behaviour may be 
looked upon as the analog of the Kelvin law for the capillary condensation 
problem. In order to determine the $a$-dependence of 
$T_{\varphi}(a)-T_{\varphi}$ one has to perform the model dependent 
numerical analysis which is postponed to future publication. Similar 
kind of qualitative analysis shows that in the opposite limit, i.e. for 
$a \rightarrow 0$ one has  $T_{\varphi}(a) \rightarrow T_{w}$. For small 
enough values of parameter $a$ the  $({\bar l}_1,\,{\bar l}_2)$ solution 
ceases to exist and only the solution $({\tilde l}_1,\,{\tilde l}_2)$ remains. 
There is no filling transition in this case. 

For a continuous transition, Fig.6b there is only one solution 
$({\bar l}_1, {\bar l}_2)$. The corresponding 
free energy contains for large $a$ the negative surface contribution 
similarly as for the first-order case discussed above. 

The above results agree qualitatively with the conclusions of our 
thermodynamic analysis in Chapter II. At $T=T_{\varphi}$ the surface 
contributions to the free energy are nonanalytical. They are given by
$\frac{2\,a}{\sin \varphi}\sigma_{w \alpha}$ for $T\,\leq\,T_{\varphi}$
and by $\frac{2\,a}{\sin \varphi}\left[\sigma_{w \alpha}\,+\,
\sigma_{\alpha \beta}(\sin \varphi\,-\,\cos \Theta)\right]$ for
$T\,>\,T_{\varphi}$, see Eq.(2.10). Now we see that also the line 
contribution which is absent in the thermodynamic analysis turns out 
to be nonanalytical at $T=T_{\varphi}(a)$. Thus to discuss the line 
contributions which determine - among others -  the 
interfacial shape one must go beyond the thermodynamic analysis. This will 
be done in Chapter IV. \\

\vspace{0.5cm}
\centerline{\bf {IIIc. The wetting transition}}
\renewcommand{\theequation}{3.\arabic{equation}} 
\vspace{0.5cm}

In order to discuss the wetting transition it is necessary to compare 
the free energy $F[{\bar f}]$ corresponding to the finite solution 
with the free energy $F_{\infty}$  corresponding to the interface 
removed infinitely far away from the substrate. This difference is equal
\begin{eqnarray}
F[{\bar f}]\,-\,F_{\infty}\,=\,\int\limits_{-a}^{a}dx\left[\frac{1}{2}
\sigma_{\alpha \beta}\left(\frac{df}{dx}\right)^2\,+\,\frac{\omega(l)}
{\sin \varphi}\right]\,\,\,,
\end{eqnarray}
where $\omega(l)=V(l)-\sigma_{w\beta}-\sigma_{\alpha \beta}$. 
It can be rewritten with the help of Eq.(3.3) as
\begin{eqnarray}
F[{\bar f}]\,-\,F_{\infty}\,=\,\frac{2\,a}{\sin\varphi}\,\omega({\bar l}_1)
\,+\nonumber \\ 2\,\sigma_{\alpha \beta}\,
\sqrt{\frac{2}{\sin\varphi}}
\int\limits_{{\bar l}_2}^{{\bar l}_1} dl \left\{\sqrt{
\sigma^{-1}_{\alpha \beta}V(l)\,-\,\sigma^{-1}_{\alpha \beta}
V({\bar l}_1)\,+\,v(\varphi)}
\,-\,\sqrt{v(\varphi)}\right\}\,\,\,.
\end{eqnarray}
In the case of the first-order potential $\omega(l)$ both contributions 
to the free energy difference in Eq.(3.7) are positive at $T_w$ which means 
that the infinite solution is the stable one. At $T_{\varphi}$ both
contributions to the free-energy difference in Eq.(3.8) 
are negative and so the finite solution is the stable one. Thus the wetting 
transition temperature $T_w(a\,\,,\,\varphi)$  is shifted from $T_w$ 
towards $T_{\varphi}$. The magnitude of this shift depends on  
parameters $a$  and $\varphi$, and the following inequality holds
\begin{eqnarray}
T_{\varphi}\,<\,T_w(a\,,\,\varphi)\,<\,T_w\,\,\,.
\end{eqnarray}
The above result - although derived for the saw-shaped 
substrate - should also hold for the first-order transitions on other 
types of periodic and weakly corrugated substrates described by the 
Hamiltonian in Eq.(3.1). 

One should also pay attention to the possibility that the first-order 
wetting transition on the planar substrate turns into critical filling 
on the corrugated substrate. This happens for "strongly" corrugated substrate 
so that the filling temperature $T_{\varphi}(a)$ lies below the spinodal 
temperature $T_{s}$ at which the potential barrier in $\omega(l)$ 
dissappears. The presence of this mechanism leading to the change of the 
order of the transition - which was pointed out in [23] in the 
context of adsorption on the wedge-shaped substrate - depends on the actual 
form of the effective potential $\omega(l)$. For certain models of 
$\omega(l)$ it may be also realized in the case of adsorption on the 
saw-shaped substrate. 

The above reasoning  is not valid for the effective potential $V(l)$ 
exhibiting the continuous transition. The continuous wetting occurs at the 
same temperature as for the planar case. \\ 
When $T\,\nearrow \,T_w$ then ${\bar l_2}$, ${\bar l_1}$, $l_{\pi}$ 
increase indefinitely and $\omega(l_{\pi}) \rightarrow 0$. 
In consequence one can neglect the potential terms in Eq.(3.4)
and derive the relation
\begin{eqnarray}
a\,\approx\,({\bar l}_1\,-\,{\bar l}_2)\tan \varphi\,\,\,.
\end{eqnarray}
This relation becomes exact at $T\,=\,T_w$. At this temperature the 
interface is flat and situated infinitely far away from the substrate. \\

\centerline{\bf {IV. Analytic results for continuous filling}}
\renewcommand{\theequation}{4.\arabic{equation}} 
\setcounter{equation}{0}
\vspace{0.5cm}

In this chapter we discuss the case when the effective potential 
$V(l)$ corresponds to the critical wetting on the planar substrate 
\begin{eqnarray}
V(l)\,=\,\sigma_{w\beta}\,+\, \sigma_{\alpha\beta}\,+ 
\,W_{0}\tau\,\exp(-l/\xi)\,+\,U_{0}\,exp(-2l/\xi)\,\,,
\end{eqnarray}
where the parameter $\tau \leq 0$ measures the distance from the planar 
substrate wetting point $\tau_{w}=0$, and the constants $W_{0}, U_{0}$ are 
positive. The solution of Eq.(3.2) has the following form 
\begin{eqnarray}
\bar{l}(x)=l_0\,+\,\xi\,\ln[1+A\,\exp(\lambda x)\,+\,B\,\exp(-\lambda x)]\,\,,
\end{eqnarray}
where the constants $l_{0}, \lambda, A, B$ are determined by 
requiring that $\bar{l}(x)$ fulfills Eq.(3.2) and the boundary conditions 
$\bar{l}'(0^{+})\,=\, \bar{l}'(a^{-})\,=\,-\cot{\varphi}$. The solution 
$\bar{l}(x)$ can be conveniently rewritten using the following 
dimensionless variables 
$y=\lambda \xi /\cot{\varphi}$, $ t=\tau /\tau_{\varphi}$,  
$ {\tau}_{\varphi}=-\sqrt{2U_{0}\sigma_{\alpha\beta} 
{W_{0}}^{-2}\sin^{-1}{\varphi}}\cos{\varphi}$, 
$ \bar{a}=a\cot{\varphi}/\xi$, 
$\chi=x\cot{\varphi}/\xi$. $\bar{a}$ measures the dimensionless 
depth of the depression at its center and $\tau_{\varphi}<0$ denotes a 
characteristic temperature for the adsorption on the saw-like substrate. For 
large opening angles considered in this paper $\tau_{\varphi}$ coincides 
with the wedge filling temperature up to the terms $0(\pi/2-\varphi)^2$. 
$t$ measures the dimensionless temperature: $t>1$ corresponds to temperature 
below the characteristic temperature $\tau_{\varphi}$ . Then 
\begin{eqnarray}
(y^2-1)\,(y^2-t^2)\,=\,t^{2}\,{\cosh}^{-2}(y\bar{a}/2)\,\,,
\end{eqnarray}
\begin{eqnarray}
\bar{l}(x)\,=\,l_{0}\,+\,\xi \ln \left[ 1\,-\,
\frac{y\cosh(y\chi)-\sinh(y\chi)}{(y^2-1)\,\sinh(y\bar{a})}\,+\,
\right. \nonumber \\ 
\left. +\, \frac{y\cosh(y(\chi-\bar{a}))-
\sinh(y(\chi-\bar{a}))}{(y^2-1)\,\sinh(y\bar{a})}\right]\,\,,
\end{eqnarray}
and 
\begin{eqnarray}
l_{0}\,=\,l_{\pi}\,-\,2\xi\ln{\left(\frac{y}{t}\right)}\,\,\,,
\end{eqnarray}
where 
$l_{\pi}\,=\,-\xi\ln\left(t\cos\varphi\sqrt{\frac{\sigma_{\alpha\beta}}
{2U_{0}\sin\varphi}}\right)$.  
The solutions of Eq.(4.3) -  which serve as the input to Eq.(4.4) - are 
parametrized by the temperature $t$ 
and the geometry of the substrate, i.e. $a$ and $\varphi$. 
The shape of the emerging interface is shown on Fig.7 for different 
temperatures. Fig.8 shows $\bar{l}_1$ and $\bar{l}_2$ as functions of 
the reduced temperature; especially pronounced is the sharp increase of 
$\bar{l}_1$ upon crossing the wedge filling temperature $t=1$. This means 
that the 
trace of the filling transition chracteristic for the adsorption in the  
wedge is also present for the saw-shaped substrate, although not in such 
a singular form as for the wedge. Also the width of the 
adsorbed layer at $x=a$, i.e. $\bar{l}_2$ does not exhibit any nonanalytic  
behavior near $t=1$; see Fig.8. A straightforward calculation shows that 
the free energy $F[\bar{f}]$  evaluated at $t=1$ tends to a constant 
for $a \rightarrow \infty$; under the same conditions the first derivative 
of $F[\bar{f}]$ with respect to the temperature grows linearly with $a$. 
This behaviour of the free energy in the limit $a \rightarrow \infty$ is 
compatible with our previous mean-field results obtained for the wedge [22]. 
On the other hand, upon approaching the planar substrate wetting 
temperature $t=0$ both $\bar{l}_{1}$ and $\bar{l}_{2}$ grow 
indefinitly which reflects the wetting transition taking place 
on this periodically corrugated substrate. 

An interesting insight into the structure of the interfacial profile 
$\bar{f}(x)$ can be obtained by 
analyzing the scaling properties of $\bar{l}(x)$ in the limit of 
large $\bar{a}$ for different temperature regimes. \\ 
It follows from Eq.(4.4) that the film thicknesses $\bar{l}_{1}$ and 
$\bar{l}_2$ are given by 
\begin{eqnarray}
\bar{l}_{1}\,=\,l_{0}\,+\, \xi \ln 
\left[\frac{y\left[y+\tanh(\frac{y\bar{a}}{2})\right]}{(y^2-1)}\right]
\end{eqnarray}
and 
\begin{eqnarray}
\bar{l}_{2}\,=\,l_{0}\,+\, \xi \ln 
\left[\frac{y\left[y-\tanh(\frac{y\bar{a}}{2})\right]}{(y^2-1)}\right]\,\,,
\end{eqnarray}
where the value of parameter $y(t, \bar{a})$ is obtained by solving Eq.(4.3). 
After substituting the relevant solutions  into Eq.(4.4) one obtains 
\begin{eqnarray}
\bar{l}_{1}\,\approx \,l_{\pi}\,+\, \left\{\begin{array}{ccl}
\xi\ln(\frac{t}{t-1}) & for & t>1 \\ 
\bar{a}\xi/2 & for & t=1 \\
\bar{a}\xi \,+\,\xi\ln(\frac{1-t^2}{2}) & for & t<1 \,\,, 
\end{array}
\right.
\end{eqnarray}
where the expressions in Eq.(4.8) hold for 
$t\bar{a}\exp(-t\bar{a}) \ll (t^2-1)$, ($t>1$), and 
$t\bar{a}\exp(-\bar{a})\ll (1-t^2)$, ($t<1$), respectively. 
Thus for $\bar{a} \rightarrow \infty$ the scaling behavior of 
$\bar{l}_{1}$ depends on the chosen temperature regime. For $t>1$ the 
value of $\bar{l}_{1}$ approaches  the width of the adsorbed layer in 
the wedge. For $t \leq 1$ the value of $\bar{l}_{1}$ 
grows linearly with $\bar{a}$; the slope of this linear increase 
is $\frac{1}{2}$ for $t=1$, and $1$ for $t < 1$. This change of 
slope is visible in Fig.9 which illustrates the scaling properties 
of $\bar{l}_1$ for different temperatures. 
  
The scaling properties of $\bar{l}_2$ are different from those of 
$\bar{l}_1$. For $t \rightarrow 0$ \, $\bar{l}_2 \rightarrow l_{\pi}$ 
and $l_{\pi}$ itself increases logarithmically. Thus for 
$t \rightarrow 0$ both $\bar{l}_1$ and $\bar{l}_2$ approach $l_{\pi}$; 
$\bar{l}_1\,-\,\bar{l}_2 \rightarrow \xi \bar{a}$, i.e. 
$\bar{f}(0)-\bar{f}(a) \rightarrow 0$. The interface becomes flat and 
the system undergoes the continuous wetting transition. \\ 

\vspace{0.5cm}
\centerline{\bf {V. The summary}}
\renewcommand{\theequation}{5.\arabic{equation}} 
\setcounter{equation}{0} 
\vspace{0.5cm}

Our main conclusion is that the complete scenario of adsorption on 
a corrugated substrate contains both the filling and the wetting 
transition. The filling transition corresponds to either partial or 
complete filling of the substrate's depressions. Then follows the 
wetting transition at which the whole substrate becomes covered by a 
macroscopically thick layer of the adsorbed phase. The detailed scenario 
of the filling transition depends on the substrate's convexity 
properties. Thermodynamic considerations show that for periodic 
substrates which are piecewise convex, i.e.  convex except for 
isolated points at which the substrate shape is nonanalytic, one has 
continuous growth of the adsorbed phase which - with increasing 
temperature - fills the substrate's depressions starting from its 
bottom and terminating at its top. The substrate's top 
position, i.e $z=b(a)$ is reached by the $\alpha$-$\beta$ interface 
at the wetting temperature $T_{w}$. On the other hand, for periodic 
substrates which are piecewise concave one has discontinuous growth.  
The $\alpha$-$\beta$ interface jumps to the position at the top 
of the substrate at the temperature  $T_{f}<T_{w}$. 
For periodic substrate shapes like $b(x)=a\,[1-\cos(\pi x/a)]$ which within 
each section contain both concave and convex parts the filling scenario 
corresponds to the jump of the interface  to the position where it makes 
contact with the convex part of the substrate. Upon further increase of 
the temperature this jump is then followed by a continuous growth of the 
interfacial position until the whole depression becomes filled. 
 
An interesting situation appears when - within each section of 
the substrate - there are more than one inflection point. This 
corresponds to varying convexity properties of the substrate which are 
then reflected in the course of the filling transition. It consists of 
jumps followed by continuous growth after which another jump comes etc. 
This highly nonuniversal scenario requires a detailed analyses which is 
postponed to future publication. 

As far as the wetting transition is concerned we argue that its order is 
the same as in the case of the corresponding planar substrate. The 
mesoscopic part of our analysis is focused on adsorption on a saw-like 
substrate which is the borderline case between piecewise convex and 
piecewise concave substrate shapes. When 
the substrate is chosen in such a way that in the planar case it corresponds 
to the critical wetting then the filling transition can be discussed to much 
extend analytically. Especially we analyze the behavior of the interfacial 
shape in the vicinity of the wedge filling temperature $T_{\varphi}$ and 
we point at a very sharp but smooth changes of the interfacial shape near 
this temperature. We also derive the scaling behavior of $\bar{l}_{1}$ with 
respect to increasing depression size for temperatures below 
and above this special temperature. 

When the substrate is chosen such that in the planar case it corresponds 
to the first-order wetting then we show that the filling transition on the 
periodically corrugated substrate is also first-order but shifted to a 
higher temperature $T_{\varphi}(a)$: $T_{\varphi} \leq T_{\varphi}(a)$. 
This scenario holds unless it is preempted by a continuous adsorption which 
may take place on the saw-shaped substrate when the filling transition 
temperature $T_{\varphi}(a)$ is shifted below the spinodal temperature 
$T_{s}$. 
We also observe that it is actually the line contribution to the free 
energy that becomes nonanalytic at $T_{\varphi}(a)$. On the contrary, 
the shift of the saw-shaped substrate wetting temperature $T_{w}(a,\varphi)$ 
with respect to the wedge wetting temperature $T_{w}$ is such that 
$T_{w}(a,\varphi) < T_{w}$. Thus one has $T_{\varphi} < T_{\varphi}(a) < 
 T_{w}(a,\varphi) < T_{w}$.  \\ 

\vspace{0.5cm}
\centerline{\bf {ACKNOWLEDGEMENTS}}
\vspace{0.5cm}

The authors gratfully acknowledge the discusssions with Siegfried Dietrich 
and Andrew Parry, and the support by the Foundation for German-Polish 
Collaboration under Grant. No. 3269/97/LN. 

\newpage
\centerline{\bf {Referrences}}
\begin{description}
\item{[1]  } {P.G. de Gennes, Rev.Mod.Phys. {\bf 57}, 827 (1985).}
\item{[2] } {S. Dietrich, in {\it Phase Transitions and Critical Phenomena}, 
edited by C. Domb and J.L. Lebowitz (Academic, London, 1988), Vol.12, p. 1.}
\item{[3]  } {M. Schick, in {\it Liquids at Interfaces,  Proceedings of the 
Les Houches Summer School in Theoretical Physics, Session XLVIII}, 
edited by J. Chavrolin, J. F. Joanny, and J. Zinn-Justin (North-Holland, 
Amsterdam, 1990), p. 415.}
\item{[4]   } {D. Andelman, J.F. Joanny, and M. O. Robbins, Europhys. Lett. 
{\bf 7}, 731 (1988); M. O. Robbins, D. Andelman, J. F. Joanny, Phys. Rev. A 
{\bf 43}, 4344 (1991); J. L. Harden and D. Andelman, Langmuir {\bf 8}, 2547 
(1992).}
\item{[5]  } {P. Pfeifer, Y.J. Wu, M.W. Cole, and J. Krim, Phys. Rev. Lett. 
{\bf 62}, 1997 (1989); P. Pfeifer, M.W. Cole, and J. Krim, Phys. 
Rev. Lett. {\bf 65}, 663 (1990).}
\item{[6]   } {M. Kardar and J.O. Indekeu, Phys. Rev. Lett. {\bf 65}, 662 
(1990); Europhys. Lett. {\bf 12}, 161 (1990); H. Li and
M. Kardar, Phys. Rev. B {\bf 42}, 6546 (1990); J. Krim and J.O. Indekeu, 
Phys. Rev. E {\bf 48}, 1576 (1993).}
\item{[7]   } {E. Cheng, M.W. Cole, and A.L. Stella, Europhys. Lett. {\bf 8}, 
537 (1989); E. Cheng, M.W. Cole, and P. Pfeifer, Phys. Rev. B {\bf 39}, 12962 
(1989).}
\item{[8]   } {G. Giugliarelli and A.L. Stella, Phys. Scripta T {\bf 35}, 34 
(1991); 
Phys. Rev. E {\bf 53}, 5035 (1996); Physica A {\bf 239}, 467 (1997); 
G. Sartoni, A.L. Stella, G. Giugliarelli, and M.R. D'Orsogna,  
Europhys. Lett. {\bf 39}, 633 (1997).}
\item{[9]} {M. Napi\'orkowski, W. Koch, and S. Dietrich, Phys. Rev. A 
{\bf 45}, 5760 (1992).}
\item{[10]  } {G. Palasantzas, Phys. Rev. B {\bf 48}, 14472 (1993); 
Phys. Rev. B {\bf 51}, 14612 (1995).}
\item{[11]  } {J.Z. Tang and J.G. Harris, J. Chem. Phys. {\bf 103}, 8201 
(1995).}
\item{[12]  } {A. Marmur, Langmuir {\bf 12}, 5704 (1996).}
\item{[13]  } {C. Borgs, J. De Coninck, R. Koteck\'y, and M. Zinque, 
Phys. Rev. Lett. {\bf 74}, 2293 (1995); K. Topolski, D. Urban, 
S. Brandon, and J. De Coninck, Phys. Rev. E {\bf 56}, 3353 (1997).}
\item{[14]  } {R. Netz and D. Andelman, Phys. Rev. E {\bf 55}, 687 (1997).}
\item{[15]  } {A.O. Parry, P.S. Swain, and J.A. Fox, J. Phys.: Condens.
Matter {\bf 8}, L659 (1996); P.S. Swain and A.O. Parry, J. Phys. A: 
Math. Gen. {\bf 30}, 4597 (1997).}
\item{[16]  } {K. Rejmer and M. Napi\'orkowski, Z. Phys. B {\bf 102}, 101 
(1997).
\item{[17]  } {P.S. Swain and A.O. Parry, Eur. Phys. J. {\bf B4}, 459 (1998).}
\item{[18]  } {S. Dietrich, in Proceedings of the NATO-ASI {\it New Approaches 
to Old and New Problems in Liquid State Theory - Inhomogeneities and Phase 
Separation in Simple,  Complex and Quantum Fluids}, Vol.{\bf C529} of NATO 
Advanced Study Institute, Messina, Italy, 1998, edited by C. Caccamo (Kluwer, 
Dordrecht,1999), p.197.}
\item{[19]  } {C. Rasc\'on, A.O. Parry, and A. Sartori}, preprint 
cond-mat/9902070} 
\item{[20]  } {Y. Pomeau, J. Coll. Interf. Sci. {\bf 113}, 5 (1985).}
\item{[21]  } {E.H. Hauge, Phys. Rev. A {\bf 46}, 4944 (1992).}
\item{[22] } {K.Rejmer, S.Dietrich, and N.Napi\'orkowski, Phys. Rev. E 
{\bf 60}, 4027 (1999).}
\item{[23]  } {A.O. Parry, C. Rasc\'on, and A.J. Wood, preprint (1999).} 
\item{[24] } {R.N. Wenzel, J. Phys. Colloid Chem. {\bf 53}, 11466 (1949); 
Ind. Eng. Chem.  {\bf 28}, 988 (1936).}
\item{[25]  } {P.S. Swain, R. Lipowsky, Langmuir {\bf 14}, 6772 (1998).}
\item{[26]  } {P. Lenz and R. Lipowsky, Phys. Rev. Lett. {\bf 80}, 1920 
(1998).}
\end{description}
 
\newpage
\centerline{\bf Figure captions}
\vspace*{5mm}

\noindent
Fig.1. Schematic configurations of the $\alpha$-$\beta$ interface at the 
depression 
extending for $x \in [-a,a]$. $x_{1}$ is the abscissa's value at which the 
interface makes contact with the substrate; $\psi$ denotes the actual 
contact angle. (a) the partial filling of the depression, (b) the interface 
located above the substrate. \\ 

\noindent
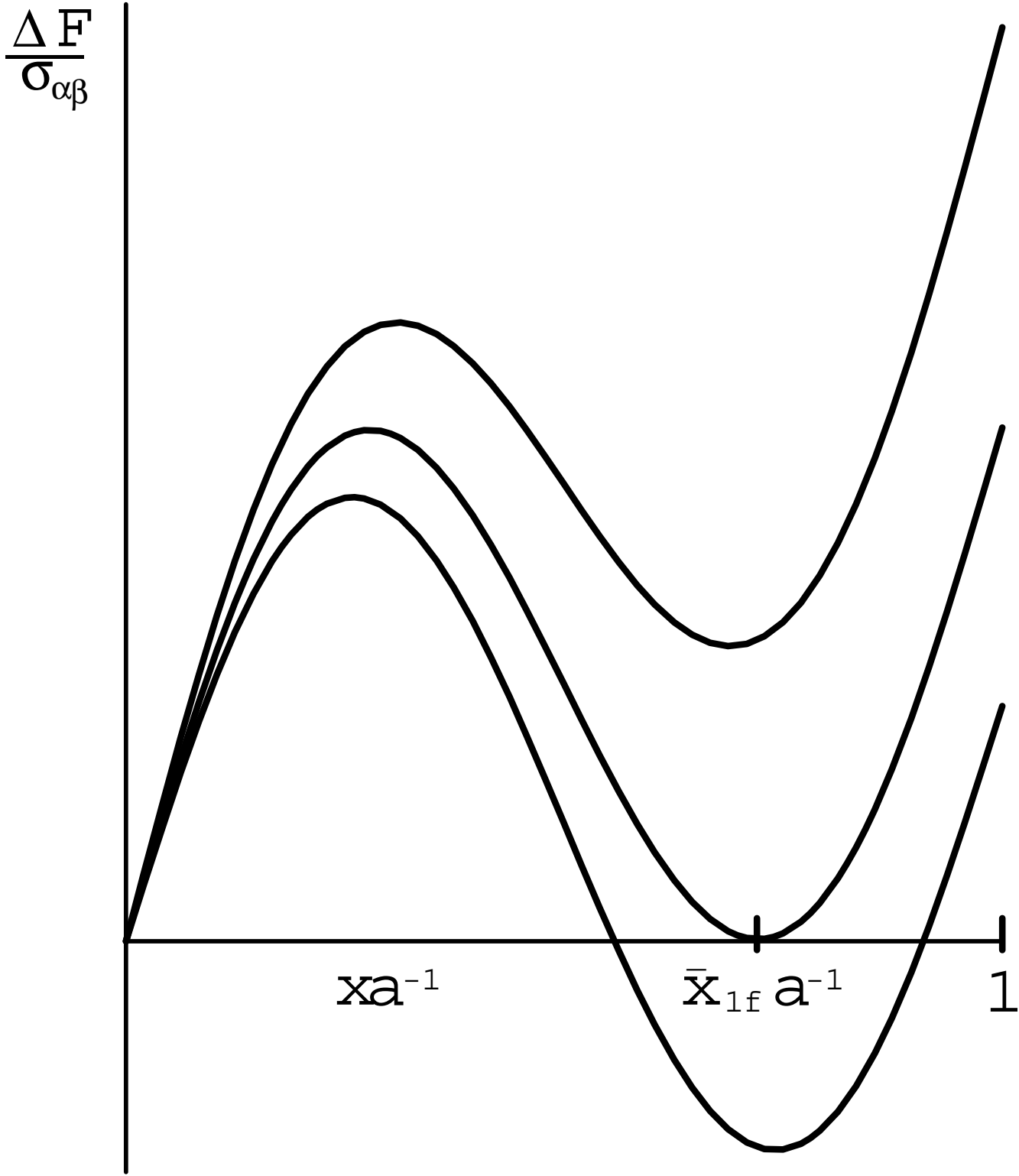
Fig.2. Plots of the excess free energy $\Delta F(x_{1})$ corresponding 
to the substrate's shape $b(x)=a(1-\cos(\pi x/a))$ for different 
values of the temperature. The curve in the middle corresponds to 
the temperature $T=T_{f}$ at which the first-order transition from the 
so-called empty depression, i.e. $\bar{x}_{1}=0$  to a partially filled 
depression with $\bar{x}_{1}=\bar{x}_{1f} < a$ takes place. \\

\noindent
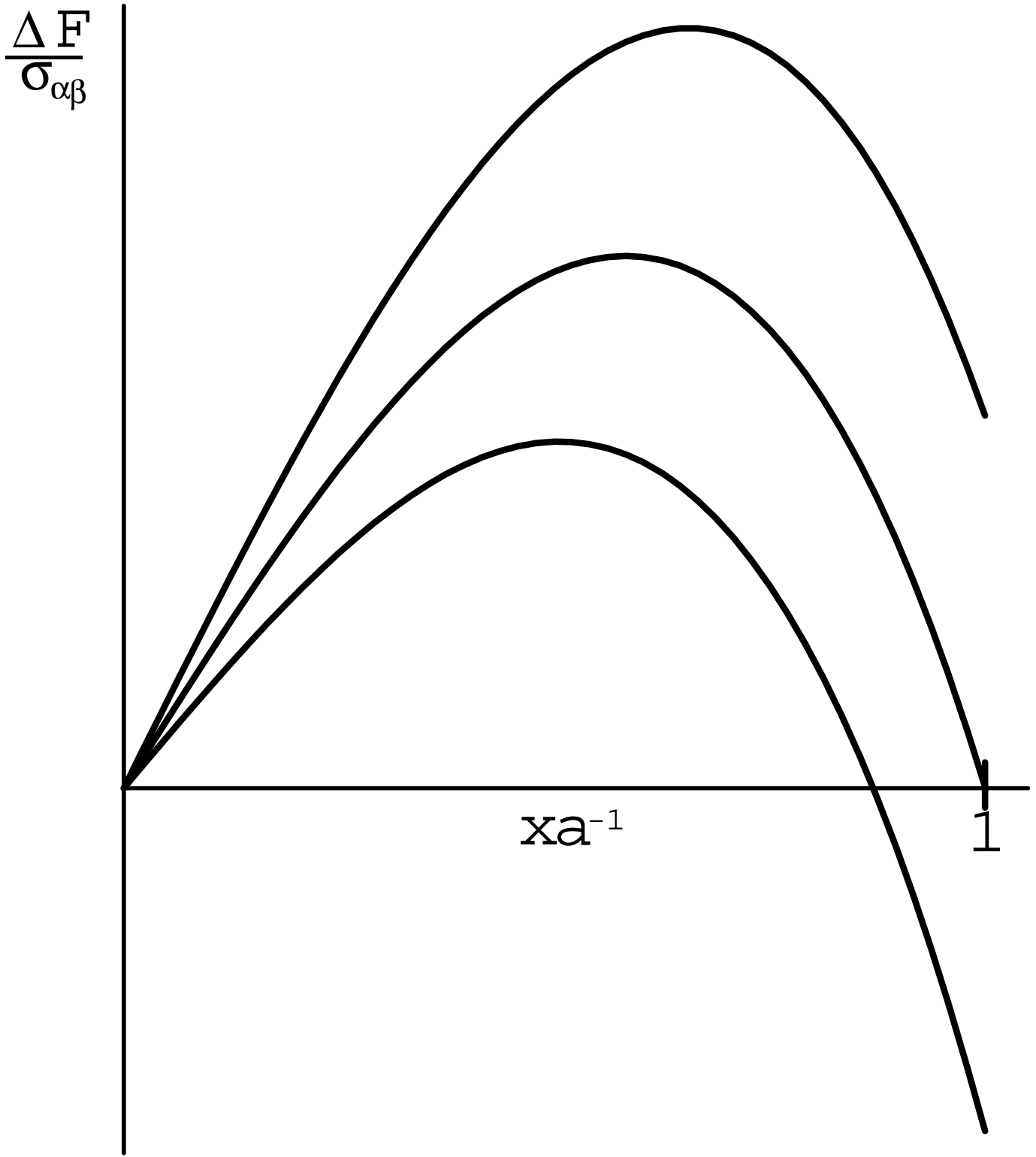
Fig.3. Schematic plots of the excess free energy $\Delta F(x_{1})$ 
corresponding to piecewise concave substrate for different temperatures. 
An example of such $\Delta F(x_{1})$ is given in Eq.(2.7). The curve in 
the middle corresponds to the temperature at which the whole depression 
becomes filled at the first-order filling transition. \\

\noindent
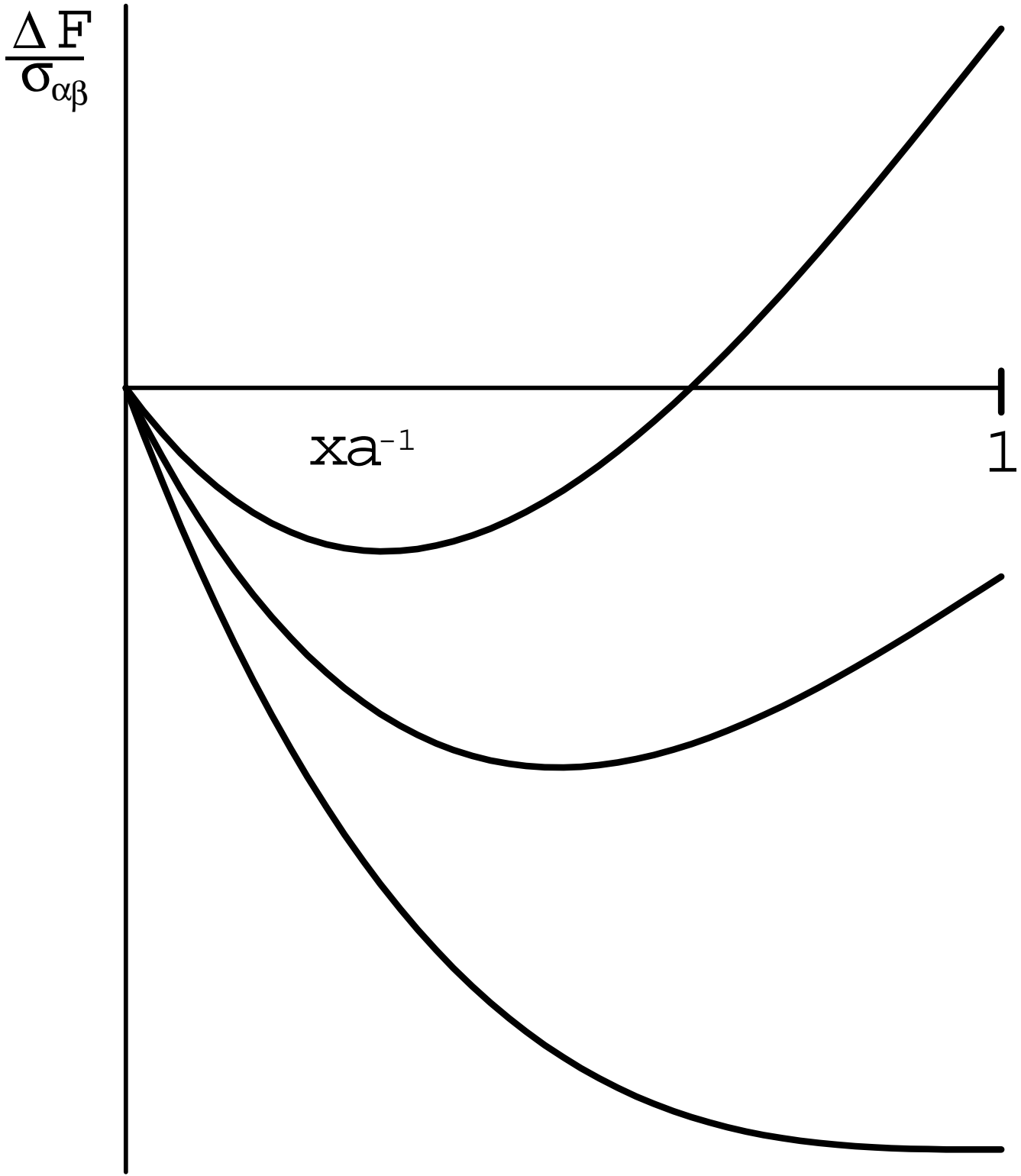
Fig.4. Schematic plots of the excess free energy $\Delta F(x_{1})$ 
corresponding to piecewise convex substrate for different temperatures. 
An example of such $\Delta F(x_{1})$ is given in Eq.(2.8). The filling of 
the depression proceeds continuously and terminates at the planar substrate 
wetting temperature $T_{w}$ at which the whole depression becomes filled. 
This corresponds to the lower curve which minimum is located at $x_{1}=a$. \\

\noindent
Fig.5. The shape of the effective interface potential 
$\sigma_{\alpha\beta}^{-1} \, \Delta V(l)$ for $T<T_{\varphi}$ in the case of 
the first-order wetting transition on the corresponding planar substrate 
(a),  
and in the case of the continuous wetting transition on the corresponding 
planar substrate (b). The dashed lines denote 
the function $v(\varphi)$, the dotted lines - $\sigma_{\alpha\beta}^{-1}\,
\Delta V_1$. $\bar{l}_{1}$ and $\bar{l}_{2}$ are the equilibrium values of 
the width of the absorbed layer at $x=0$ and $x=a$, respectively. 
$1-\cos\theta$ is the limiting value of 
$\sigma_{\alpha\beta}^{-1} \, \Delta V(l)$ for $l \rightarrow 
\infty$ and is denoted by thin dashed lines. 
$\sigma_{\alpha\beta}^{-1} \, \Delta V(l_{\pi})=0$; the value of 
$l_{\pi}$ which gives the equilibrium width of the wetting layer on the 
corresponding planar substrate is not marked on this figure. The dotted line 
approaches the dashed line for $T \nearrow T_{\varphi}$. \\ 

\noindent
Fig.6.  The shape of the effective interface potential 
$\sigma_{\alpha\beta}^{-1} \, \Delta V(l)$ for $T>T_{\varphi}$ in the case of 
the first-order wetting transition on the corresponding planar substrate 
(a),  
and in the case of the continuous wetting transition on the corresponding 
planar substrate (b). The notation is the same as in Fig.5. In the case of 
the first-order wetting there exist two solutions described in the text 
as $(\tilde{l}_{1},\tilde{l}_{2})$ and $(\bar{l}_{1},\bar{l}_{2})$. For 
clarity reasons only $\tilde{l}_{1}$ is marked on Fig.6a; $\tilde{l}_{2}$ 
practically coincides with $\bar{l}_{2}$. For the same reasons we have not 
plotted separately horizontal dotted lines corresponding to 
$\Delta \tilde{V}_{1}$ and $\Delta V_{1}$ although they 
differ in values. For $T \searrow T_{\varphi}$ the dotted line approaches 
the asymptote $1-\cos\theta$ while $\bar{l}_{1} \rightarrow \infty$. \\ 

\noindent
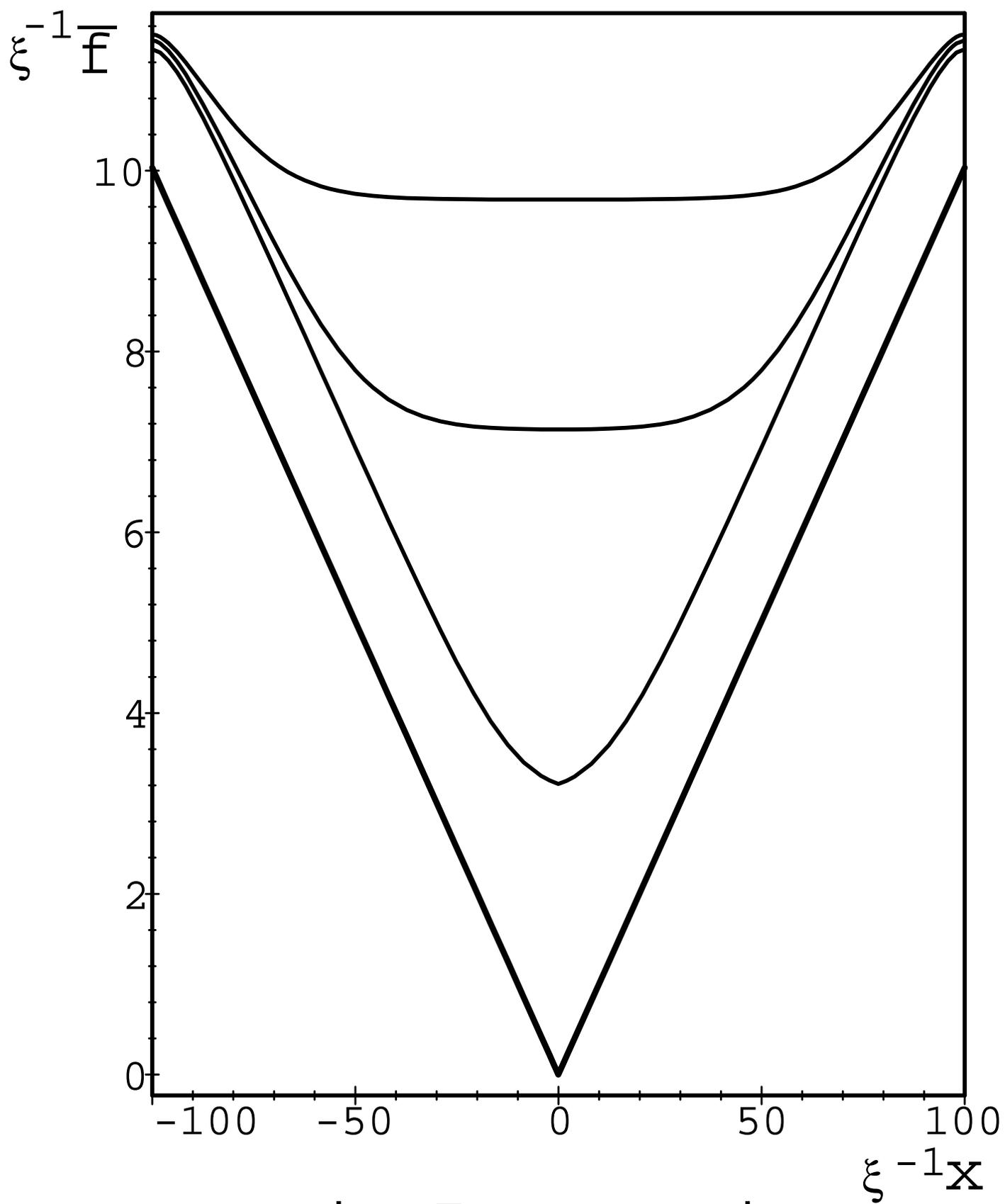
Fig.7. The shapes of the $\alpha$-$\beta$ interface for different 
dimensionless temperatures approaching the temperature $t=1$ which 
corresponds to the critical filling transition on the saw-shaped substrate. 
The three curves, in increasing order, correspond to the temperatures 
$t=1.2$, $t=1$, and $t=0.9$, respectively. \\

\noindent
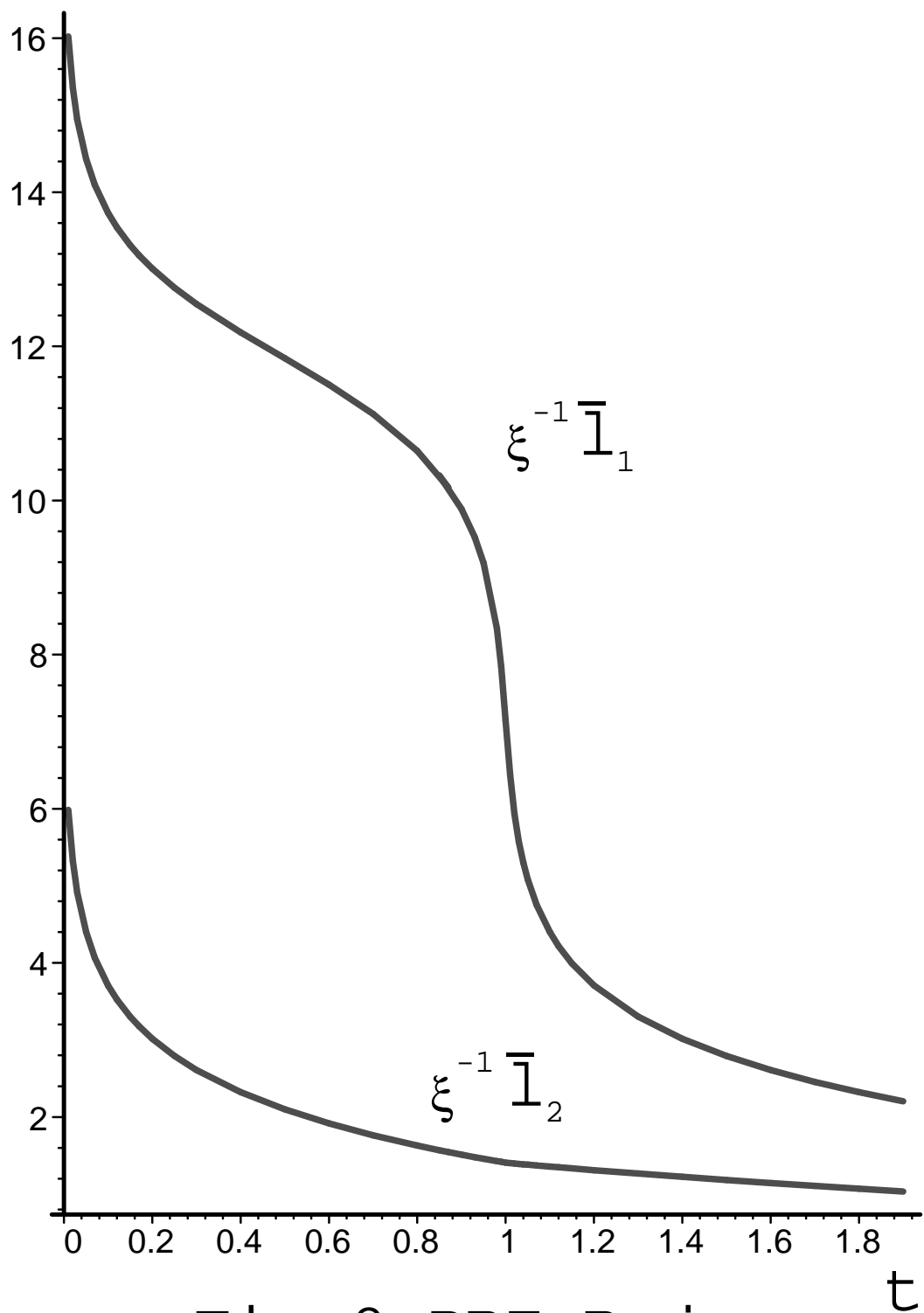
Fig.8. The widths of the adsorbed layer $\bar{l}_1$ and $\bar{l}_2$ as 
functions of the dimensionless temperature $t$. For $t \searrow 0$ 
both  $\bar{l}_1$ and $\bar{l}_2$ tend to $\infty$ with their difference 
kept constant and equal to $a$; on this figure $\bar{a}=10$. \\

\noindent
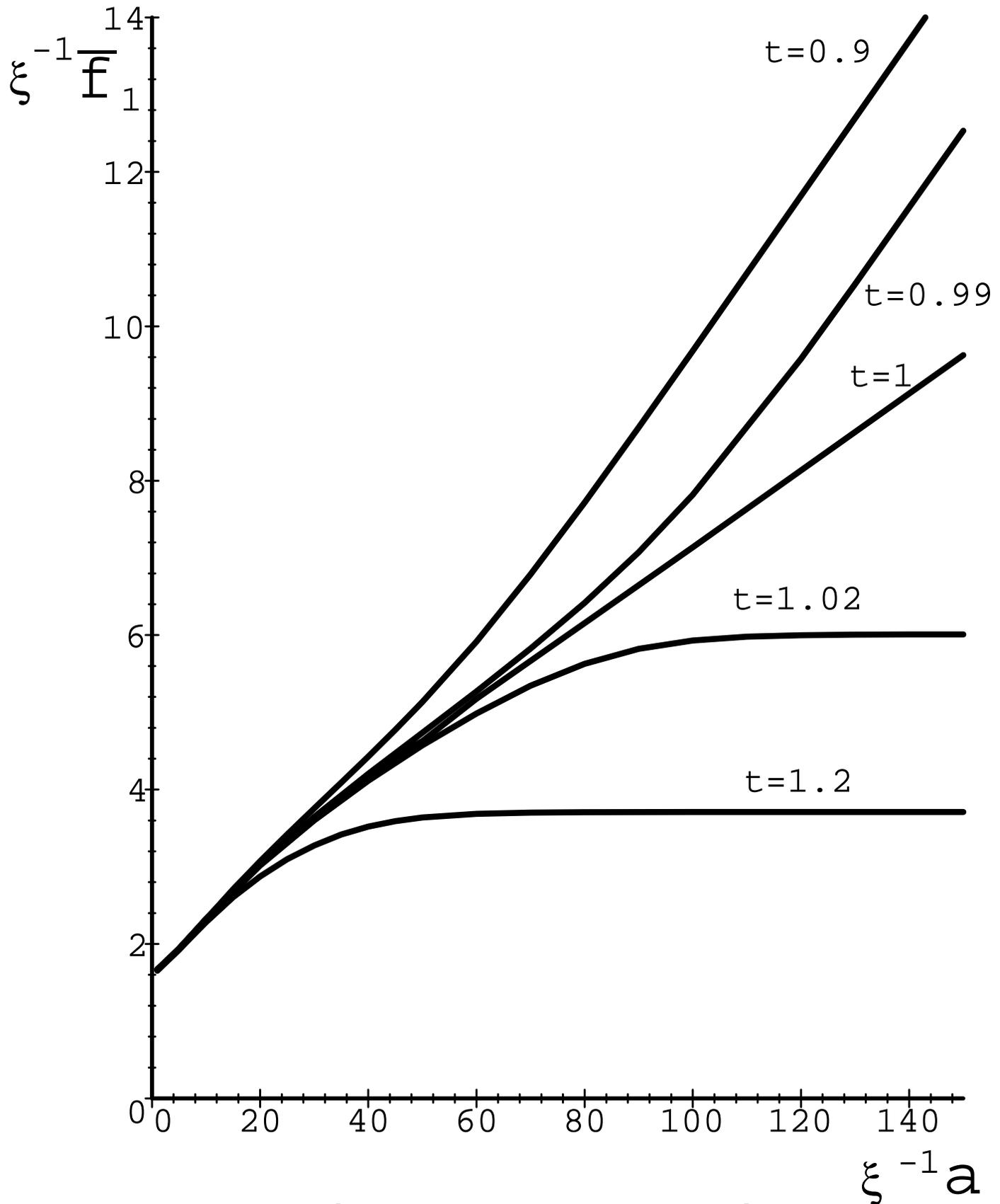
Fig.9. The scaling behaviour of $\bar{l}_1=\bar{f}(0)$ as function of $a$ 
for different temperatures in the vicinity of $t=1$. For $t=1$  
$\bar{l}_1 \sim \bar{a}/2$, for $t<1$ $\bar{l}_1 \sim \bar{a}$. For 
$t>1$ $\bar{l}_1$ approaches the value corresponding to the adsorption 
in the wedge. 

\end{document}